# THE EFFECT OF THE DIGIT SLICING ARCHITECTURE ON THE FFT BUTTERFLY

*Yazan Samir, Rozita Teymourzadeh IEEE member*

Department of Electrical and Electronic, Faculty of Engineering,
Institute of Microengineering and Nanoelectronics, Blok Inovasi 2,
University Kebangsaan Malaysia, 43600 Bangi, Selangor, Malaysia
Email: yazansamir@yahoo.com, rozita_teymourzadeh@yahoo.com

## ABSTRACT

*Most communications systems tend to achieve bandwidth, power and cost efficiencies to capable to describe modulation scheme. Hence for signal modulation orthogonal frequency division multiplexing (OFDM) transceiver is introduced to cover communications demand in four generation. However high performance Fast Fourier Transforms (FFT) as a main heart of OFDM acts beyond the view. In order to achieve capable FFT, design and realization of its efficient internal structure is key issues of this research work. In this paper implementation of high performance butterfly for FFT by applying digit slicing technique is presented. The proposed design focused on the trade-off between the speed and active silicon area for the chip implementation. The new architecture was investigated and simulated with the MATLAB software. The Verilog HDL code in Xilinx ISE environment was derived to describe the FFT Butterfly functionality and was downloaded to Virtex II FPGA board.*

## 1. INTRODUCTION

With the increasing use of transceiver in communication system the usage of and requirement for signal processing has increased. Among the digital signal processing block FFT is the most critical processor to transfer one function into another, which is called the frequency domain representation, or simply the Discrete Fourier Transform (DFT) of the original time domain. DFT is the main and important procedure in the data analysis, system design and implementation [1]. In order to reduce the complexity computation of the FFT algorithm many modules have been designed and implemented in different platforms. These modules focus on radix order or twiddle factor to perform a simply and efficient algorithm which includes the higher radix FFT [2], the mixed-radix FFT [3], the prime-factor FFT [4], the recursive FFT [5], low-memory reference FFT [6], Multiplier-less based FFT [7, 8, and 9] and application-specific integrated circuits (ASIC) system such as [10]. ASIC-based system can fit real application for low-power or high performance; however, it is very solid to modify the function [11]. The study of the digit slicing technique has been dealt by [12, 13, and 14] for the digital filters. The design and implementation of Digit slicing FFT has been discussed one time before in [15]. This paper proposed the similar idea with [15] by using a different algorithm and different platform which help to improve the performance and get higher speed. Recently, FPGAs has become an applicable option to direct hardware solution performance in the real time application. In this paper, digit slicing architecture was proposed to design the digit-slicing butterfly. Which is a portion of the computation that combines the results of smaller discrete Fourier transforms into a larger DFT. The FFT butterfly multiplication is the most crucial part in causing the delay in the computation of the FFT. In view of the fact, the twiddle factors in the FFT processor was known in advance hence we proposed to use the digit slicing multiplier to replace with conventional multiplier in FFT butterfly.

The paper structure was organized as follow; section 2 describes the FFT architecture in brief, whilst section 3 explains the butterfly conventional architecture. The digit slicing architecture will be discussed in section 4 followed by section 5 that proposes the design of digit slicing butterfly architecture in detail. Finally section 6 and 7 shows implementation result and conclusion respectively.

## 2. FAST FOURIER TRANSFORM (FFT)

A useful method to transform domain from the time domain to the frequency domain and the reverse for the implementation on digital hardware is the discrete Fourier transform (DFT). For *N*-point DFT of a complex data sequence x (n) is defined in equation (1).

$$X(k)=\sum_{n=0}^{N-1}x(n)W_N^{kn} \quad ,k=0,1,.......,N-1 \qquad (1)$$

Where *x(n)* and *X(k)* are complex numbers, and $W_N^{kn}=e^{-j2\pi/N}$ is the twiddle factor. The DFT of N-point finite sequence represents harmonically related frequency components of $x(n)$. The direct computation of equation (1) requires the order of $N^2$ operations where $N$ is the transform size. In 1965, Cooley and Tukey have found the new technique to reduce the order of complexity operations of DFT from $N^2$ to ($Nlog_2N$). Consequently, a huge number of FFT algorithms have been developed such as Radix-2, radix-4 and split radix algorithms. These algorithms mostly used for practical applications due to their simple structure and constant butterfly geometry. In general, higher-radix FFT algorithm has fewer numbers of complex multiplications whereas radix-2 FFT algorithm is the simplest form in all FFT algorithms. Furthermore, it has a regularity that makes it suitable for VLSI implementation as shown in the fallowing equation (2).

$$X[m] = \sum_{n=0}^{\frac{N}{2}-1} x[2n] W_{\frac{N}{2}}^{nm} + W_N^m \sum_{n=0}^{\frac{N}{2}-1} x[2n+1] W_{\frac{N}{2}}^{nm} \qquad (2)$$

FFT algorithm relies on a divide and conquers methodology, which divides the $N$ coefficient points into smaller blocks in different stages. The first stage computes with groups of two coefficients, yielding $N/2$ blocks, each computing the addition and subtraction of the coefficients scaled by the corresponding twiddle factors, called a butterfly for its cross-over appearance. These results are used to compute the next state of $N/4$ blocks, which will then combine the results of two previous blocks, combining 4 coefficients at this point. This process is repeated until we have one main block, with a final computation of all $N$ coefficients [9].

## 3. CONVENTIONAL BUTTERFLY ARCHITECTURE

The conventional radix-2 DIT butterfly architecture is consisting of complex data I/O, complex multiplier and finally complex adder and subtractor Fig. 1.

Consider $A$ and $B$ are the complex input data, the complex twiddle factor considered as $W = Wr - jWi$.

Finally the complex output are $X$ and $Y$. The index r and i represent the real and imaginary parts respectively.

$$X = A + BW_N^r \qquad (3)$$

$$Y = A - BW_N^r \qquad (4)$$

$$(Xr + jXi) = (Ar + jAi) + [(Wr + jWi) \times (Br + jBi)] \qquad (5)$$

$$(Yr + jYi) = (Ar + jAi) - [(Wr + jWi) \times (Br + jBi)] \qquad (6)$$

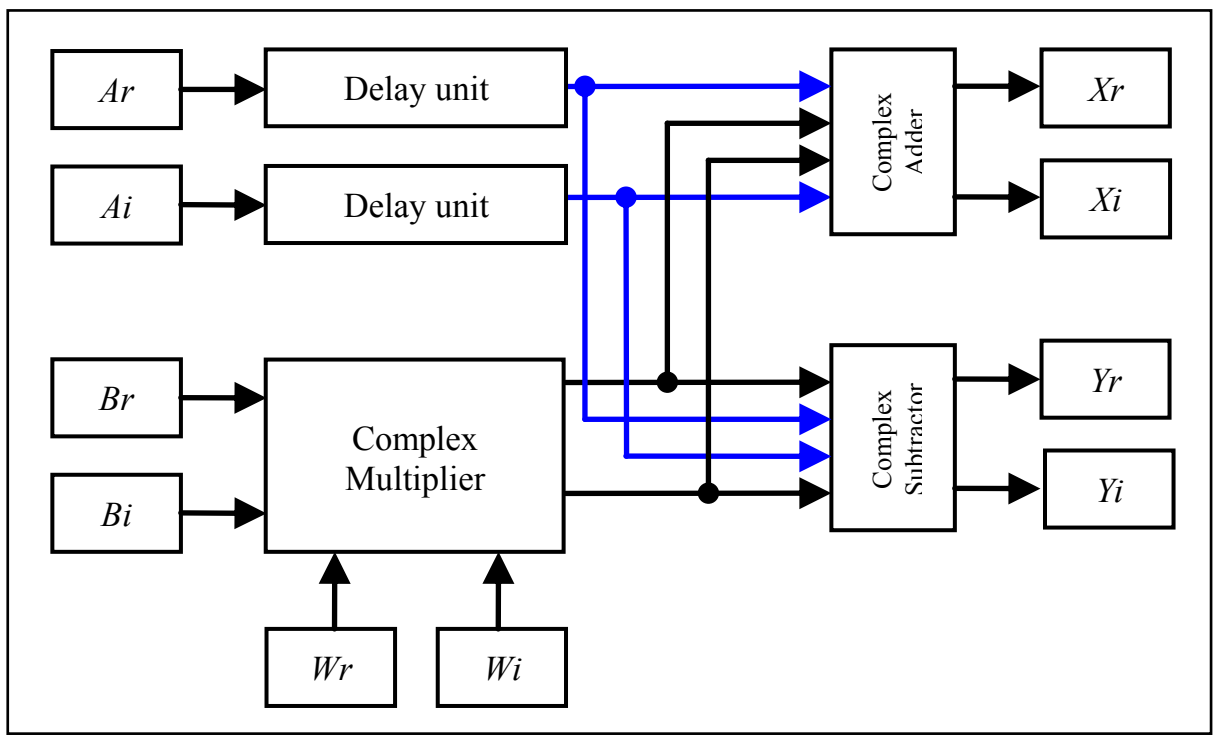

**Figure 1.** Radix-2 DIT FFT Butterfly Architecture

The implementation of the complex multiplier required four real multipliers and two real adders Fig. 2. The complex multiplier was determined equation (11).

$$(Br + jBi) \times (Wr + jWi) = (Br \times Wr) - (Bi \times Wi)] + [(Br \times jWi) + (jBi \times Wr)] \qquad (7)$$

The real and imaginary parts of the multiplication result is $[(Br \times Wr) - (Bi \times Wi)]$ and $[(Br \times jWi) + (jBi \times Wr)]$ respectively.

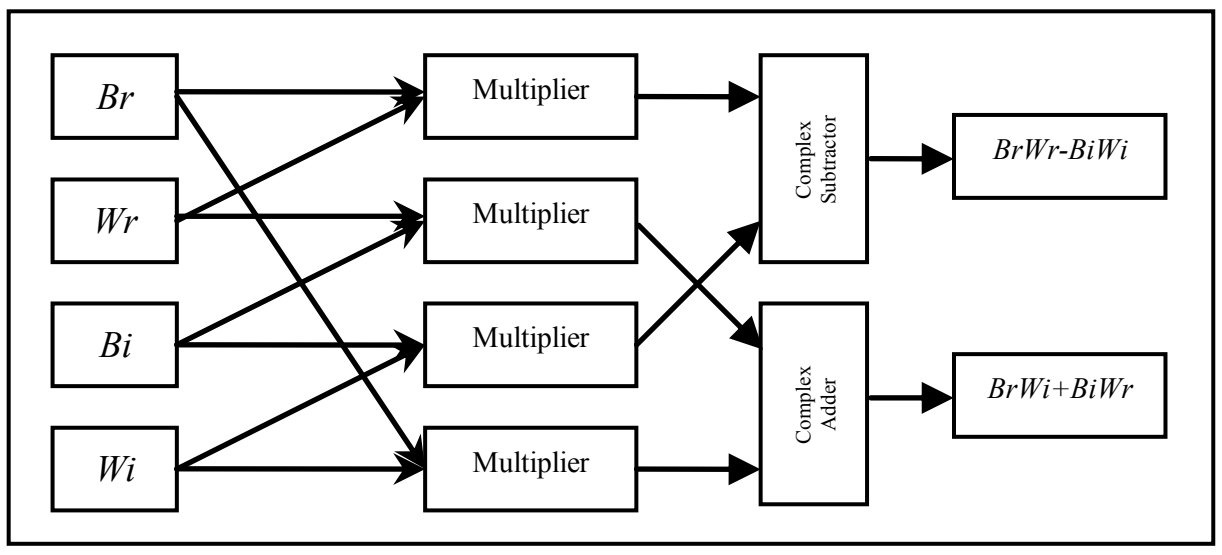

**Figure 2.** The Complex Multiplier Structure.

The complex adder is required two real adders to perform addition functionality.

$$(Ar + jAi) + (Br + jBi) = (Ar + Br) + j(Ai + Bi) \qquad (8)$$

## 4. DIGIT SLICING ARCHITECTURE

The concept behind the digit slicing architecture is any binary number can be sliced into a few blocks of shorter binary numbers, with each block carrying a different weight. In this paper, the fixed-point 2's complements arithmetic has been chosen to represent the input data, which are singed numbers with absolute value less than one. The absolute value of the input data $x$ with length of B bits ($x^0, x^1, x^2, \ldots, x^{B-1}$) has been represented in 2's complement as:

$$x = \sum_{k=0}^{B-1} 2^{-j} x^j \qquad (9)$$

To represent the sliced data, there are many different algorithms. Depend on the data type and word length, different structures can be introduced. In this paper, the fundamental sliced algorithm will be presented as following:

$$x = \left[ \sum_{k=0}^{b-1} 2^{pk} X_k \right] 2^{-(pb-1)} \qquad (10)$$

Where $x$ is sliced into $b$ blocks and $p$ is bit widths per block.

$$X_k = \sum_{j=0}^{p-1} 2^j X_{k,j} \qquad (11)$$

Where $X_{k,j}$ are all either ones or zeros except for $X_{k=b-1,j=p-1}$ which is zero or minus one. This algorithm (6 and 7) applies when the sliced data word length is $2^2$ such as 4, 8, 16, ect. bits. Another algorithm to represent the sliced data with word length $2^2+1$ such as 5, 9, 17, ect. bits can be dealt as following:

$$x = \sum_{k=0}^{p-1} \left[2^p\right]^{-k} X_k \qquad (12)$$

Where $x$ is a decimal number whose absolute value is less than one, and is sliced into b blocks each of p bits wide. The most significant block is k = 0 and this contains only the sign bit of $x$ plus leading dummy zeros to make up a block of length p bits [15].

$$X_{k=0} = 0 \;\; or \;\; -1 \;\; only$$

$$X_k = \sum_{j=0}^{p-1} 2^j X_{k,j} \; ; \; X_{k,j} = 0 \; or \; 1 \; only \;\; for \; k \neq 0. \quad (13)$$

As a comparison between the first and the second example, the second algorithm required one extra block to deal with the sign bit only this makes the design more complicated and requires more hardware for the implementation. In this paper, the first digit-slicing algorithm has been chosen to build the digit-slicing FFT butterfly structure. Therefore, any complex numbers, F, can be sliced into smaller blocks b, each having a shorter word length, p, as illustrated in following equations:

$$F = F_R + j\,F_I \quad (14)$$

$$F = \left[\sum_{k=0}^{b-1} 2^{pk} F_{Rk}\right] 2^{-(pb-1)} + j\left[\sum_{k=0}^{b-1} 2^{pk} F_{Ik}\right] 2^{-(pb-1)} \quad (15)$$

$$F_{Rk} = \sum_{j=0}^{p-1} 2^j F_{Rk,j} \quad and \quad F_{Ik} = \sum_{j=0}^{p-1} 2^j F_{Ik,j} \quad (16)$$

Where $F_{Ik,i}$ and $F_{Rk,I}$ have values which are either zero or one.

## 5. DIGIT SLICING BUTTERFLY ARCHITECTURE

The novel butterfly architecture was designed and investigating accordingly. In order to reduce the complexity computation and enhanced the throughput, digit slicing butterfly obtained by applying the digit slicing technique. As mentioned in section 3 the butterfly structure contains of one complex multiplier, one complex adder and one complex subtractor.

The digit slicing architecture has been applied for the butterfly input to slice the data to four groups each carrying four bits as shown in Fig. 3.

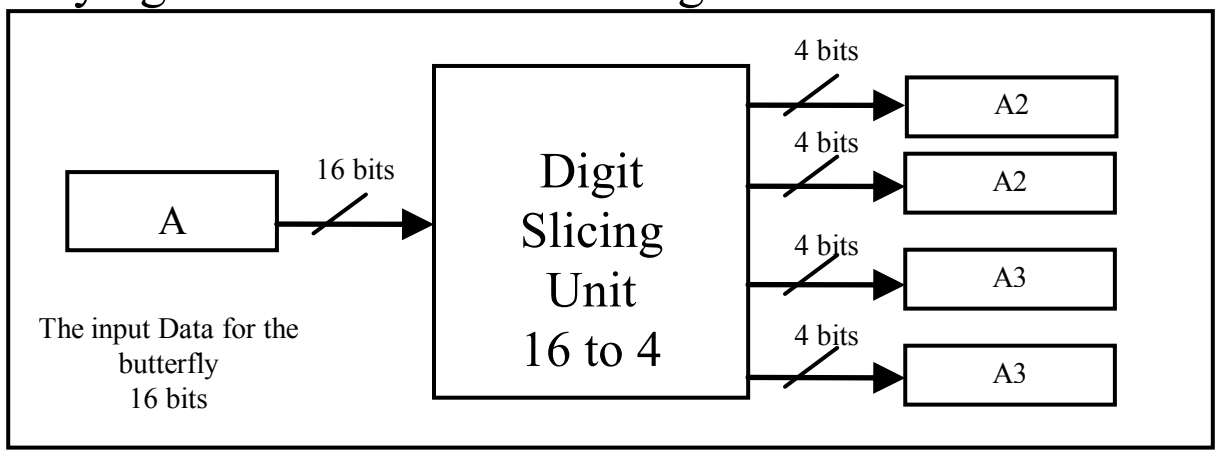

**Figure 3.** Digit Slicing Structure

The multiplication regarded as the most important operations for most signal processing systems, but it complex and expensive operation. Many techniques have been introduced for reducing the size and improving the speed of multipliers. Since the twiddle factor in FFT processor are known in advanced a special design of digit slicing multiplier has been proposed to perform the multiplication with the twiddle factor. The digit slicing architecture has been applied for the input data and sliced to four groups each has four bits to be multiplied by the constant $W = Wr - jWi$ 16 bits 2's complement fixed point, which has absolute value less than one to represent the FFT twiddle factor.

By applying equation (10 and 11) to equation (5 and 6) the digit slicing butterfly output will be:

$$X = (Xr + jXi) \quad and \quad Y = (Yr + jYi)$$

$$A = (Ar + jAi) \quad and \quad B = (Br + jBi)$$

By slicing the input data A and B into b blocks each carrying p bits wide.

$$A = \left[\sum_{k=0}^{b-1} 2^{pk} A_k\right] 2^{-(pb-1)} \; where \; A_k = \sum_{j=0}^{p-1} 2^j A_{k,j}$$

Where $A_{k,j}$ are all either ones or zeros except for $A_{k=b-1,j=p-1}$ which is zero or minus one.

The same for the input B

$$B = \left[\sum_{k=0}^{b-1} 2^{pk} B_k\right] 2^{-(pb-1)} \; where \; B_k = \sum_{j=0}^{p-1} 2^j B_{k,j}$$

Where $B_{k,j}$ are all either ones or zeros except for $B_{k=b-1,j=p-1}$ which is zero or minus one. The output X has been defined as:

$$X = \left[\sum_{k=0}^{b-1} 2^{pk} X_k\right] 2^{-(pb-1)} \; where X_k = \sum_{j=0}^{p-1} 2^j X_{k,j}$$

Where $X_{k,j}$ are all either ones or zeros except for $X_{k=b-1,j=p-1}$ which is zero or minus one.

Putting all the slicing equations above into equation (3)

$$X = A + BW_N^r$$

$$\left[\sum_{k=0}^{b-1} 2^{pk} X_k\right] 2^{-(pb-1)} = \left[\sum_{k=0}^{b-1} 2^{pk} A_k\right] 2^{-(pb-1)} + \left[\sum_{k=0}^{b-1} 2^{pk} B_k\right] 2^{-(pb-1)} \times W_N^r$$

$$X_k = A_k + B_k \times W_N^r \quad (17)$$

$$X_k \; is \; complex \; number \quad X_k = X_{rk} + jX_{ik}$$

$$\text{Re}\,al \; part \; of \; X_{rk} = A_k + B_k \times W_N^r \quad and \quad \text{Im}\,ag \; part \; of \; X_{ik} = -B_k \times W_N^r$$

Same step for output X has been applied to get output Y

$$\left[\sum_{k=0}^{b-1} 2^{pk} Y_k\right] 2^{-(pb-1)} = \left[\sum_{k=0}^{b-1} 2^{pk} A_k\right] 2^{-(pb-1)} - \left[\sum_{k=0}^{b-1} 2^{pk} B_k\right] 2^{-(pb-1)} \times W_N^r$$

$$Y_k = A_k - B_k \times W_N^r \quad (18)$$

$$Y_k \; is \; complex \; number \quad Y_k = Y_{rk} + jY_{ik}$$

$$\text{Re}\,al \; part \; of \; Y_{rk} = A_k - B_k \times W_N^r \quad and \quad \text{Im}\,ag \; part \; of \; Y_{ik} = B_k \times W_N^r$$

Finally, the complex output has been represented as following:

$$X_{rk} = A_{rk} + B_{rk} \times Wr + B_{ik} \times Wi \quad (19)$$

$$X_{ik} = A_{ik} + B_{ik} \times Wr - B_{rk} \times Wi \quad (20)$$

$$Y_{rk} = A_{rk} - B_{rk} \times Wr - B_{ik} \times Wi \quad (21)$$

$$Y_{ik} = A_{ik} - B_{ik} \times Wr + B_{rk} \times Wi \quad (22)$$

The constant twiddle factor was stored in look-up table ROM.

## 6. IMPLEMENTATION RESULT

Two different modules have been implemented for radix-2 DIT butterfly. The first module uses the conventional architecture for the butterfly where the twiddle factors are stored in ROM and called by the butterfly to be multiplied with the inputs using the dedicated high speed multiplier equipped with the Virtex-II FPGA and the other module uses the digit slicing multiplier-less architecture to perform the multiplication with the twiddle factor. Both modules has been built and tested in MATLAB Fig. 4, then coded in Verilog and synthesized using the XST - Xilinx Synthesis Technology tool. The target FPGA was Xilinx Virtex-II XC2V500-6-FG456 FPGA [17]. ModelSim simulation result of digit slicing radix-2 DIT butterfly is shown in Fig. 5, while the synthesis results for the two models are presented in Table 1, which demonstrates the hardware specifications for the design. It shows the maximum throughput of 535.90 MHz for the mentioned system. As well as the digit slicing multiplier has performed the maximum throughput of 609 MHz.

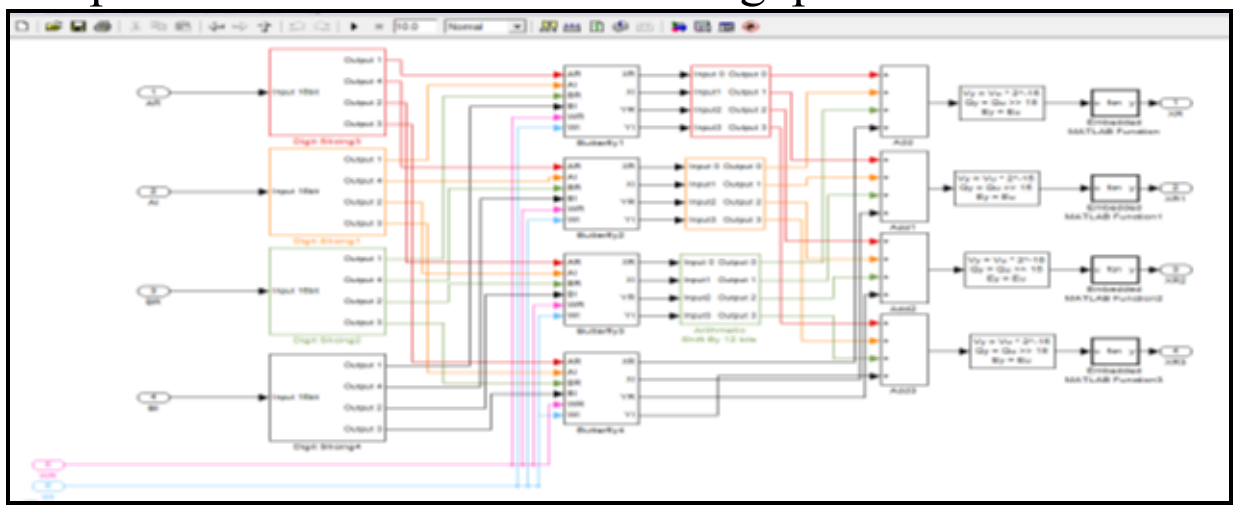

**Figure 4.** MATLAB design of Digit Slicing Butterfly

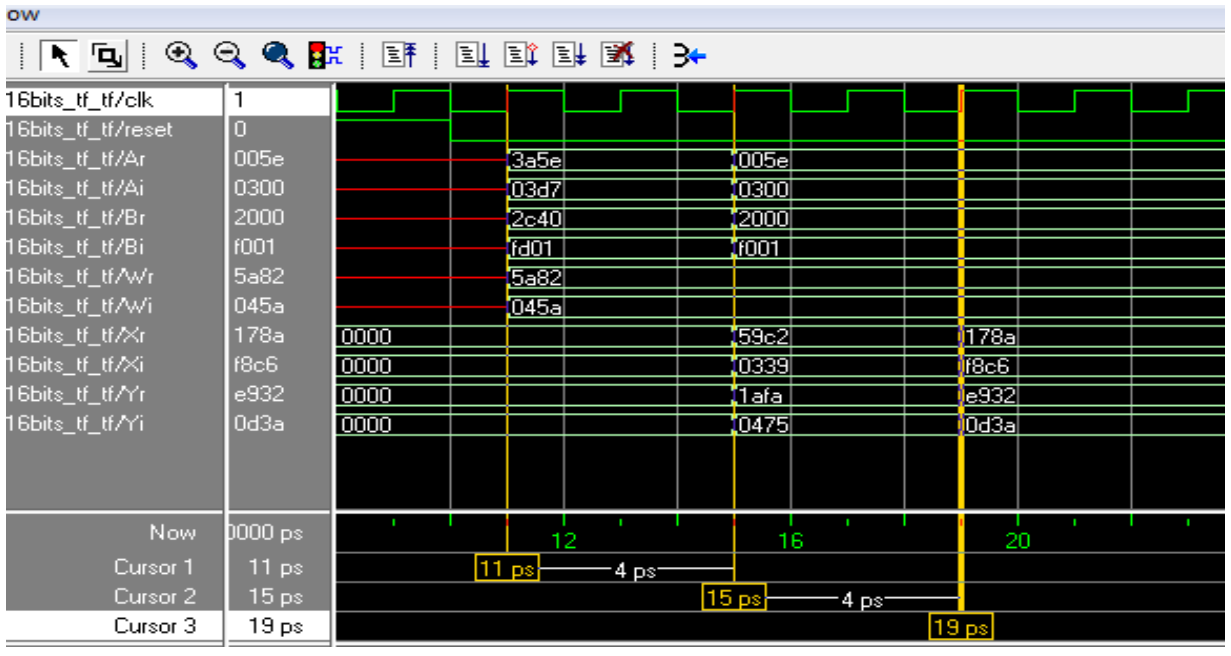

**Figure 5.** Simulation result of digit slicing Butterfly

TABLE 1.
HARDWARE SPECIFICATIONS OF THE DIGIT-SLICING BUTTERFLY

| Xilinx Virtax-II FPGA XC2v250-6FG456 | Total equivalent gate | Maximum Frequency MHz |
|---|---|---|
| Conventional butterfly | 18,408 | 198.987 |
| Digit-Slicing Butterfly | 31,159 | 535.90 |
| Conventional 16 bits Multiplier | 4,131 | 220.16 |
| Digit-Slicing 16 bits Multiplier | 6,483 | 609.60 |

## 7. CONCLUSION

This paper presented design and implementation of digit slicing butterfly for FFT structure. The implementation has been coded in Verilog hardware descriptive language and was tested on Xilinx Virtex-I1 XC2V500-6- FG456 prototyping FPGA board. A maximum clock frequency of 535.90 MHz has been obtained from the synthesis report for the digit slicing butterfly.